\documentclass[twocolumn]{aastex6}

\shorttitle{Habitability in the Omega Centauri Cluster}
\shortauthors{Stephen R. Kane \& Sarah J. Deveny}

\begin{document}

\title{Habitability in the Omega Centauri Cluster}

\author{
  Stephen R. Kane\altaffilmark{1},
  Sarah J. Deveny\altaffilmark{2}
}
\altaffiltext{1}{Department of Earth Sciences, University of
  California, Riverside, CA 92521, USA}
\altaffiltext{2}{Department of Physics \& Astronomy, San Francisco
  State University, 1600 Holloway Avenue, San Francisco, CA 94132,
  USA}
\email{skane@ucr.edu}


\begin{abstract}

The search for exoplanets has encompassed a broad range of stellar
environments, from single stars in the solar neighborhood to multiple
stars and various open clusters. The stellar environment has a
profound effect on planet formation and stability evolution and is
thus a key component of exoplanetary studies. Dense stellar
environments, such as those found in globular clusters, provide
particularly strong constraints on sustainability of habitable
planetary conditions. Here, we use Hubble Space Telescope observations
of the core of the Omega Centauri cluster to derive fundamental
parameters for the core stars. These parameters are used to calculate
the extent of the Habitable Zone of the observed stars. We describe
the distribution of Habitable Zones in the cluster and compare them
with the stellar density and expected stellar encounter rate and
cluster dynamics. We thus determine the effect of the stellar
environment within the Omega Centauri core on the habitability of
planets that reside within the cluster. Our results show that the
distribution of Habitable Zone outer boundaries generally lie within
0.5~AU of the host stars, but that this small cross-sectional area is
counter-balanced by a relatively high rate of stellar close encounters
that would disrupt planetary orbits within the Habitable Zone of
typical Omega Centauri stars.

\end{abstract}

\keywords{astrobiology -- planetary systems -- stars: kinematics and
  dynamics -- globular clusters: individual (Omega Centauri)}


\section{Introduction}
\label{intro}

Thus far, searches for exoplanets have primarily occurred around field
stars, such as the exoplanet survey undertaken by the {\it Kepler}
mission \citep{bor10}. The prospect of exoplanet detection in globular
cluster environments is particularly enticing since they represent a
relatively old stellar population and allow studies of how cluster
dynamics influences planet formation and evolution
\citep{fre06,sok07,spu09,dej12,por15,cai17}. A survey for transiting
exoplanets among lower main-sequence (MS) stars in the globular
cluster NGC~6397 by \citet{nas12} did not detect any significant
exoplanet signatures. The primary target of exoplanet searches in
globular clusters has been 47 Tucanae (47~Tuc). Observations of 34,000
stars in the 47~Tuc core by \citet{gil00} using the Hubble Space
Telescope (HST) did not detect any transiting planets, despite
predictions of almost 20 planet detections. Follow-up grand-based
observations by \citet{wel05} in the uncrowded outer regions of 47~Tuc
also did not detect transiting planets, indicating that the apparent
lack of planets in the core may not be solely due to cluster
dynamics. However, a recalculation by \citet{mas17} of the expected
planet occurrence rates in 47~Tuc based on {\it Kepler} results
determined that only a handful of planets detections should be
expected, thus potentially reducing the statistical significance of
the initial null result.

Omega Centauri ($\omega$~Cen, NGC~5139) is a globular cluster that is
also the possible remnant of a disrupted dwarf galaxy
\citep{gne02,noy08}. As the largest globular cluster in the Milky Way
galaxy, $\omega$~Cen provides an ideal stellar population for
investigations concerning the interaction of radiation environments
and stellar dynamics \citep{mer97,rei06,van06}. An additional
advantage of studying cluster stars in the context of exoplanets is
that they tend to have measured luminosities that enable the
calculation of the Habitable Zone (HZ) for each of the stars
\citep{kop13,kop14}. Such calculations in turn allow for the
quantification of habitability within these dense cluster environments
and thus direct the motivation of terrestrial exoplanet searches in
globular clusters.

Here we present an analysis of {\it HST} observations of the core of
$\omega$~Cen and a calculation of HZs for the observed stars. In
Section~\ref{hst} we outline the {\it HST} observations, including the
calibration, passbands, and quantity of stars. Section~\ref{stellar}
describes the methodology used to select MS stars and the derivation
of stellar parameters. Section~\ref{hz} presents the calculations of
the HZ for the stellar sample and discusses their distribution. The
convolution of the HZ boundaries and the stellar dynamics is addressed
in Section~\ref{dynamics}, taking into account the mean distance
between stars and the rate of close stellar
encounters. Section~\ref{implications} discusses the implications of
the HZ calculations for potential habitability within $\omega$~Cen and
how this is balanced by planetary orbit disruptions from the close
encounter rate between stars. We finally provide concluding remarks in
Section~\ref{conclusions}.


\section{HST Observations of $\omega$ Centauri}
\label{hst}

The $\omega$~Cen cluster has been extensively observed by {\it HST}
with a variety of science goals using, for example, the Advanced
Camera for Surveys Wide Field Channel (ACS/WFC) \citep{coo02}. Many
groups have performed a wide range of studies that include its
multiple stellar populations \citep{mil17}, proper motion
\citep{bel18}, optical counterparts to X-ray sources \citep{coo13} and
the search for the possible intermediate mass black hole
\citep{noy08,and10,hag13}.

The data used for this project was taken from the recently published
photometric catalog from \citet{bel17a} which is now the newest and
most extensive photometric analysis of $\omega$~Cen ever
undertaken. The {\it HST} data includes 26 Wide Field Camera 3
(WFC3/UVIS) filters (18 WFC3/UVIS and 11 WFC3/IR), of which we
selected two wide-band filters, F438W and F555W, with $34 \times 350$s
and $27 \times 40$s exposures, respectively. We chose the specific
F438W and F555W filters because they are the most comparable to the
standard Johnson $B$ and $V$ filters, respectively. They cover roughly
a $5\arcmin \times 5\arcmin$ field-of-view (FOV) that contains the
$2.37\arcmin$ radius core of $\omega$~Cen with over 470,000 stars.


\section{Derivation of Stellar Parameters}
\label{stellar}

\begin{figure}
  \includegraphics[clip,width=8.5cm]{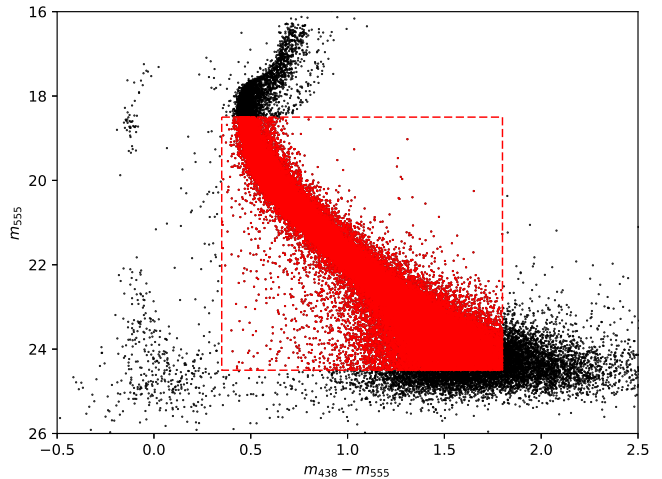}
  \caption{The CMD of the $\omega$~Cen core, including {\it HST}
    photometry of $\sim$410,000 stars. The red stars inside the dashed
    box are those selected as MS stars and results in a sample size of
    $\sim$350,000 stars.}
  \label{cmd}
\end{figure}

The initial stellar sample consisted of 470,000 stars from which we
aimed to extract a sample consisting of MS stars in the core. Due to
the nature of photometric uncertainties in the crowded core of
$\omega$~Cen, we implemented a cut on the stellar sample that was a
simple reduction of the null rms error values for the calibrated
magnitudes that reduced the list by approximately 12,000 stars. The
second step in revising the stellar sample was a cut to only include
the core of $\omega$~Cen because we wanted to use a region of
relatively uniform stellar density. This was a significant reduction
that left us with a total of just over 410,000 stars. Finally, we
wanted to isolate the MS because habitable zones around giant branch
stars or stellar remnants were not what we wanted. This cut was done
by using a color-magnitude diagram (CMD) to plot the MS, where the
passbands are the F438W and F555W filters described in
Section~\ref{hst}. We then manually created boundaries around the MS
and selected only the stars that are inside the boundaries. The
resulting CMD and MS selection boundaries are shown in
Figure~\ref{cmd}. The specific cuts for the MS selection are $0.3 \leq
m_{438} - m_{555} \leq 1.8$ and $18.5 \leq m_{555} \leq 24.5$. Note
that many of the stars at the extreme tail-end of the MS are excluded
from our selection due to large error values on the measured F438W
and/or F555W magnitudes in that region. After all three of the above
cuts were applied, we were left with a total sample size of
$\sim$350,000 stars.

\begin{figure*}
  \begin{center}
    \begin{tabular}{cc}
      \includegraphics[clip,width=8.2cm]{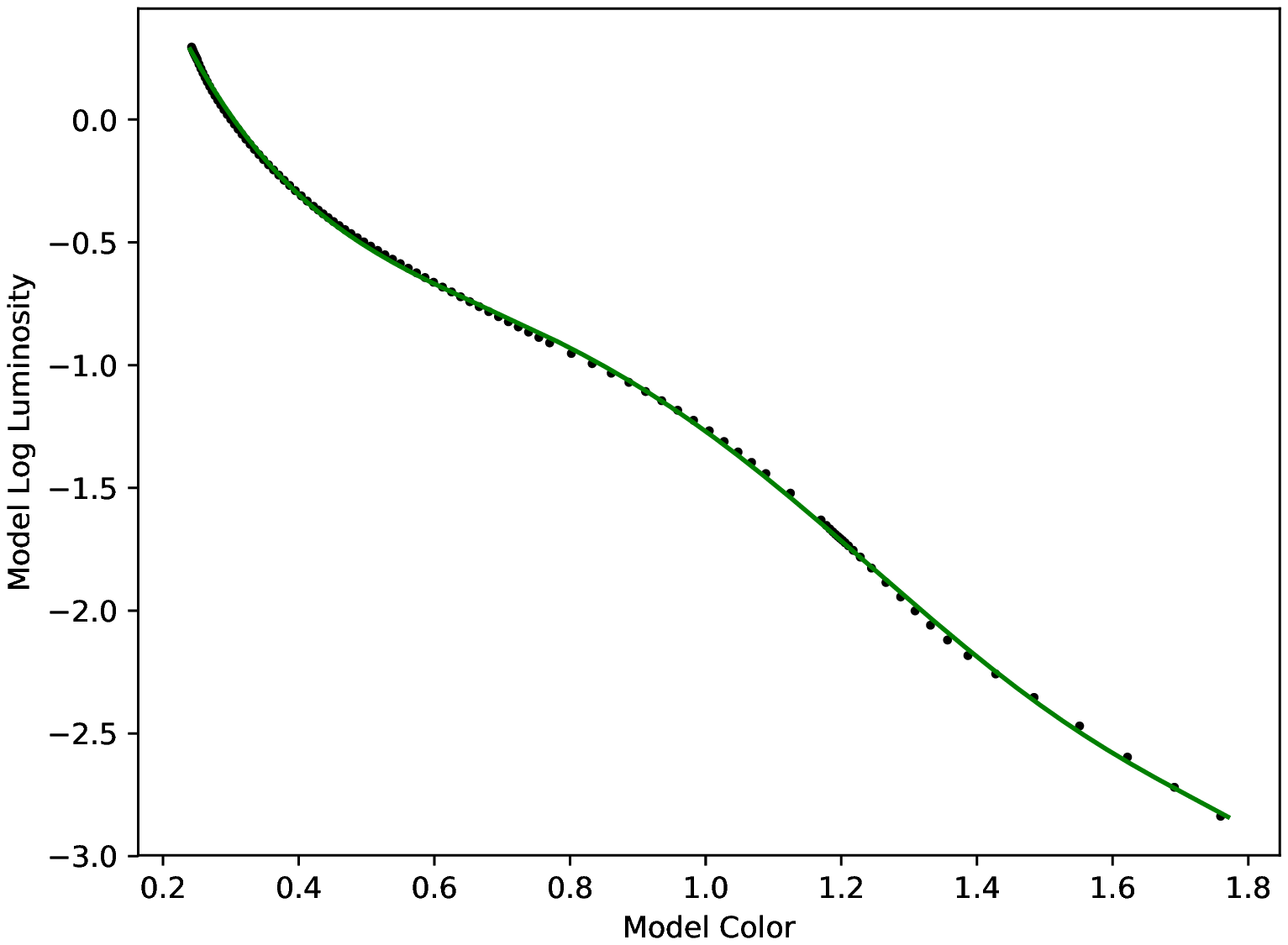} &
      \includegraphics[clip,width=8.2cm]{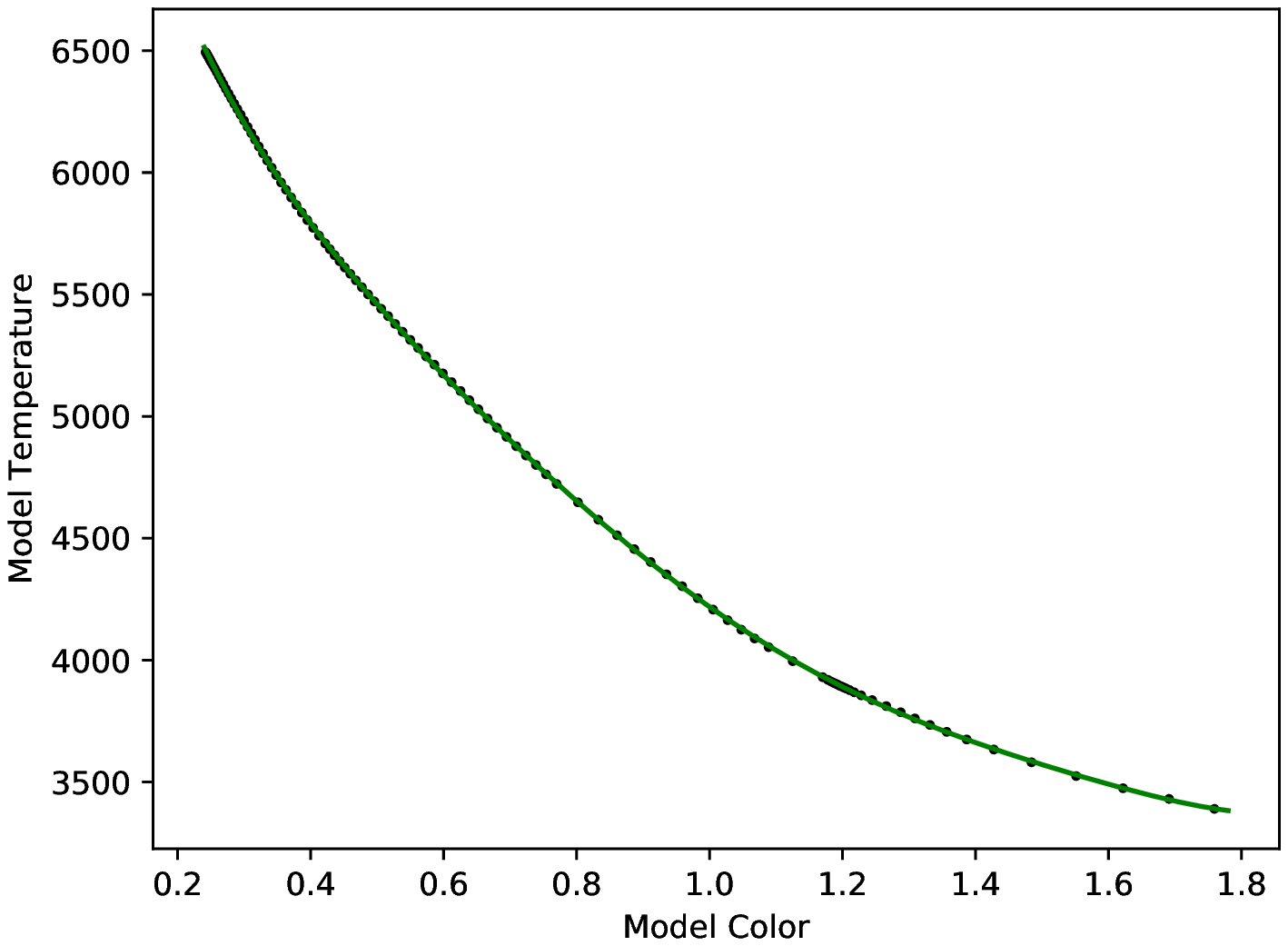}
    \end{tabular}
  \end{center}
  \caption{The isochrone model luminosity (left) and effective
    temperature (right) as a function of the $B-V$ model color. The
    green solid line shows the result of a sixth-order polynomial fit
    to the isochrone data, enabling the measurement of luminosity and
    effective temperature values for the $\omega$~Cen stars observed
    with {\it HST}.}
  \label{models}
\end{figure*}

For the subsequent analysis, we required the luminosity ($L_\star$)
and effective temperatures ($T_\mathrm{eff}$) of the $\omega$~Cen
stars. To calculate these stellar parameters, we used an isochrone
model of $\omega$~Cen based on the WFC3/UVIS filter system from the
``Dartmouth Stellar Evolution Database'' \citep{dot08}. The
$\omega$~Cen core MS stars have been demonstrated to exhibit a large
range of metallicities. \citet{bel17b} described three populations of
stars consisting of bMS ($\mathrm{[Fe/H]} \sim -1.4$), rMS
($\mathrm{[Fe/H]} \sim -1.7$), and MSe ($\mathrm{[Fe/H]} \sim -0.7$)
stars, where the bMS and rMS stars comprise $\sim$65\% of the core
population. We tested the effects of this metallicity diversity on our
subsequent HZ analysis and found that metallicities within the range
of rMS to MSe stars have a negligible effect on the HZ
calculations. We therefore selected a metallicity of $\mathrm{[Fe/H]}
= -1.49$ \citep{vil14} since that lies within the metallicity
distribution for the bulk of the $\omega$~Cen core stars. We used the
11.5~Gyr age isochrone model, consistent with the age of 11.52~Gyr for
the age of $\omega$~Cen \citep{for10}. We then performed a
least-squares sixth-order polynomial fit between the $B-V$ colors and
the $L_\star$ and $T_\mathrm{eff}$ isochrone model values. The
isochrone data and their associated fits (green lines) are shown in
Figure~\ref{models}, where the left panel shows $L_\star$ vs $B-V$
color, and the right panel shows $T_\mathrm{eff}$ versus $B-V$
color. These derived relationships allowed us to calculate the
$L_\star$ and $T_\mathrm{eff}$ values, along with propagated
uncertainties, from the colors measured from the $\omega$~Cen HST
observations (see Figure~\ref{cmd}). We also included the effects of
reddening toward $\omega$~Cen in these calculations using the results
of the multiband photometry of \citet{cal17}. In our sample,
$\sim$80\% of the stars have luminosities less than 25\% of solar
luminosity, consistent with the relatively large amount of low-mass
stars shown in Figure~\ref{cmd} and also with the aged population of
the $\omega$~Cen stars.


\section{Habitable Zones in $\omega$ Centauri}
\label{hz}

The HZ is generally defined as the region around a star where a
terrestrial planet may possibly have surface conditions suitable for
liquid water, given sufficient atmospheric pressure. The extent of
this region has been quantified by a number of sources, most
prominently by \citet{kas93} and further revised by
\citet{kop13,kop14}. The primary boundaries that are used to describe
the HZ are the conservative HZ (CHZ) that consider theoretical
calculations of maintaining temperate surface conditions, and the
optimistic HZ (OHZ) that uses empirically derived assumptions
regarding the prevalence of surface water on Venus and Mars
\citep{kas14,kan16}. The CHZ and OHZ boundaries were used, for
example, to create a catalog of Kepler HZ planets by \citet{kan16} and
are represented graphically for known exoplanetary systems in the
Habitable Zone Gallery\footnote{\tt http://hzgallery.org}
\citep{kan12}.

\begin{deluxetable*}{ccccccc}
  \tablewidth{0pc}
  \tablecaption{\label{tab} Measured and derived values for a sample
    of $\omega$~Cen stars.}
  \tablehead{
    \colhead{$B-V$} &
    \colhead{$L_\star$ ($L_\odot$)} &
    \colhead{$T_\mathrm{eff}$ (K)} &
    \colhead{OHZ$_i$ (AU)} &
    \colhead{CHZ$_i$ (AU)} &
    \colhead{CHZ$_o$ (AU)} &
    \colhead{OHZ$_o$ (AU)}
  }
  \startdata
$ 0.438\pm 0.002$ & $ 0.413\pm 0.005$ & $5659\pm   8$ & $ 0.486\pm 0.003$ & $ 0.615\pm 0.004$ & $ 1.089\pm 0.007$ & $ 1.149\pm 0.007$ \\
$ 0.633\pm 0.004$ & $ 0.191\pm 0.005$ & $5079\pm  11$ & $ 0.341\pm 0.005$ & $ 0.432\pm 0.006$ & $ 0.780\pm 0.011$ & $ 0.823\pm 0.011$ \\
$ 0.493\pm 0.022$ & $ 0.324\pm 0.045$ & $5481\pm  69$ & $ 0.435\pm 0.030$ & $ 0.551\pm 0.038$ & $ 0.979\pm 0.068$ & $ 1.033\pm 0.071$ \\
$ 0.905\pm 0.047$ & $ 0.081\pm 0.076$ & $4413\pm 102$ & $ 0.228\pm 0.108$ & $ 0.288\pm 0.136$ & $ 0.535\pm 0.253$ & $ 0.565\pm 0.267$ \\
$ 0.686\pm 0.066$ & $ 0.161\pm 0.078$ & $4938\pm 171$ & $ 0.315\pm 0.076$ & $ 0.399\pm 0.097$ & $ 0.724\pm 0.176$ & $ 0.764\pm 0.185$ \\
$ 0.608\pm 0.056$ & $ 0.209\pm 0.077$ & $5147\pm 153$ & $ 0.355\pm 0.065$ & $ 0.450\pm 0.083$ & $ 0.810\pm 0.149$ & $ 0.854\pm 0.158$ \\
$ 1.270\pm 0.087$ & $ 0.013\pm 0.211$ & $3801\pm 102$ & $ 0.092\pm 0.759$ & $ 0.117\pm 0.961$ & $ 0.223\pm 1.840$ & $ 0.236\pm 1.941$ \\
$ 1.180\pm 0.024$ & $ 0.022\pm 0.059$ & $3918\pm  34$ & $ 0.120\pm 0.164$ & $ 0.152\pm 0.207$ & $ 0.289\pm 0.394$ & $ 0.305\pm 0.416$ \\
$ 0.605\pm 0.076$ & $ 0.211\pm 0.106$ & $5155\pm 209$ & $ 0.357\pm 0.089$ & $ 0.452\pm 0.113$ & $ 0.813\pm 0.204$ & $ 0.858\pm 0.215$ \\
$ 0.579\pm 0.042$ & $ 0.232\pm 0.062$ & $5228\pm 116$ & $ 0.373\pm 0.050$ & $ 0.472\pm 0.063$ & $ 0.847\pm 0.114$ & $ 0.893\pm 0.120$ \\
$ 0.699\pm 0.040$ & $ 0.154\pm 0.048$ & $4904\pm 104$ & $ 0.309\pm 0.048$ & $ 0.392\pm 0.060$ & $ 0.712\pm 0.110$ & $ 0.751\pm 0.116$ \\
$ 0.499\pm 0.078$ & $ 0.316\pm 0.153$ & $5463\pm 238$ & $ 0.430\pm 0.104$ & $ 0.544\pm 0.132$ & $ 0.969\pm 0.236$ & $ 1.022\pm 0.249$ \\
$ 0.788\pm 0.082$ & $ 0.118\pm 0.103$ & $4682\pm 199$ & $ 0.272\pm 0.119$ & $ 0.345\pm 0.151$ & $ 0.633\pm 0.277$ & $ 0.667\pm 0.292$ \\
$ 0.502\pm 0.037$ & $ 0.312\pm 0.072$ & $5454\pm 112$ & $ 0.427\pm 0.049$ & $ 0.541\pm 0.062$ & $ 0.963\pm 0.111$ & $ 1.016\pm 0.117$ \\
$ 0.864\pm 0.027$ & $ 0.093\pm 0.040$ & $4504\pm  61$ & $ 0.243\pm 0.052$ & $ 0.308\pm 0.066$ & $ 0.570\pm 0.122$ & $ 0.601\pm 0.129$ \\
$ 0.953\pm 0.022$ & $ 0.067\pm 0.041$ & $4312\pm  46$ & $ 0.209\pm 0.063$ & $ 0.264\pm 0.080$ & $ 0.493\pm 0.150$ & $ 0.520\pm 0.158$ \\
$ 0.904\pm 0.041$ & $ 0.081\pm 0.067$ & $4415\pm  90$ & $ 0.228\pm 0.094$ & $ 0.289\pm 0.119$ & $ 0.536\pm 0.221$ & $ 0.565\pm 0.234$ \\
$ 0.446\pm 0.022$ & $ 0.398\pm 0.051$ & $5632\pm  73$ & $ 0.478\pm 0.031$ & $ 0.605\pm 0.039$ & $ 1.071\pm 0.069$ & $ 1.130\pm 0.073$ \\
$ 0.339\pm 0.008$ & $ 0.727\pm 0.022$ & $6030\pm  33$ & $ 0.630\pm 0.010$ & $ 0.798\pm 0.012$ & $ 1.399\pm 0.021$ & $ 1.475\pm 0.023$ \\
$ 0.711\pm 0.050$ & $ 0.149\pm 0.058$ & $4874\pm 127$ & $ 0.304\pm 0.060$ & $ 0.385\pm 0.075$ & $ 0.700\pm 0.137$ & $ 0.739\pm 0.145$
  \enddata
\end{deluxetable*}

\begin{figure*}
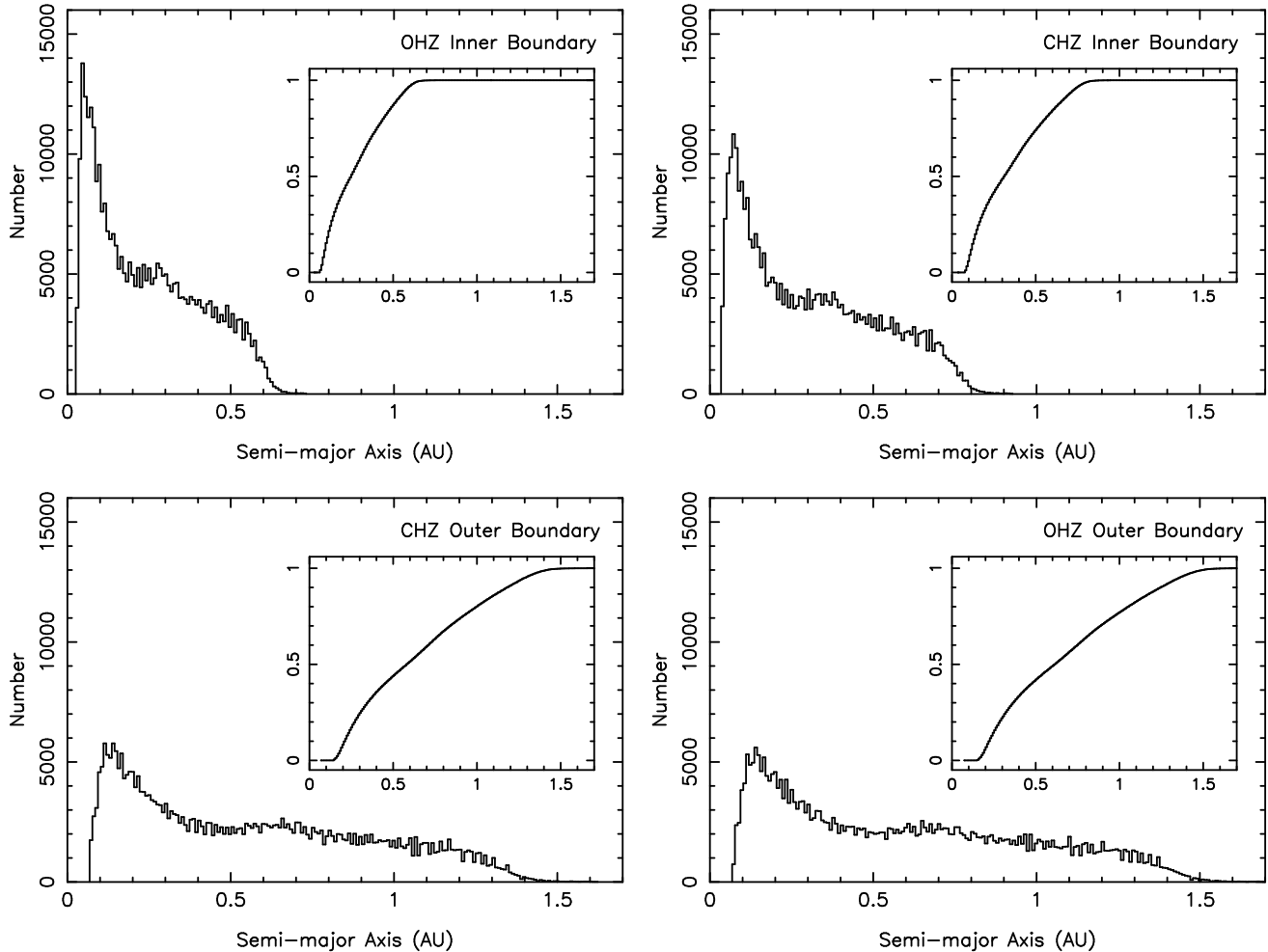

  \begin{center}
    \begin{tabular}{cc}
      \includegraphics[angle=270,width=8.5cm]{f03a.ps} &
      \includegraphics[angle=270,width=8.5cm]{f03b.ps} \\
      \includegraphics[angle=270,width=8.5cm]{f03c.ps} &
      \includegraphics[angle=270,width=8.5cm]{f03d.ps}
    \end{tabular}
  \end{center}
  \caption{Histograms of the HZ boundaries for the inner OHZ
    (top-left), inner CHZ (top-right), outer CHZ (bottom-left), and
    outer OHZ (bottom-right). All four panels use the same axis scales
    for ease of comparison. The inset panel is a normalized cumulative
    histogram, where the x-axis is also semi-major axis. The
    distribution of HZ boundaries matches the distribution of stellar
    properties derived in Section~\ref{stellar}, where the sample is
    dominated by low-mass stars.}
  \label{histfig}
\end{figure*}

Using the stellar properties for $\omega$~Cen derived in
Section~\ref{stellar}, the HZ relationships found in \citet{kop14},
and the HZ error propagation methodology from \citet{cha16}, we
calculated the CHZ and OHZ boundaries for each of the stars in our
sample. As shown by \citet{kan14a}, the stellar parameter
uncertainties can have a significant impact on the determination of HZ
boundaries, thus the need to include the appropriate error propagation
in our analysis. A sample of the complete table for our stellar
parameters and HZ calculations is shown in Table~\ref{tab}, where the
subscripts of $i$ and $o$ are used for the inner and outer HZ
boundaries respectively. The distribution of each of the HZ boundaries
is represented in Figure~\ref{histfig}. The panels of the figure are
fixed to identical scales for ease of comparison. The inset panel
shows a normalized cumulative histogram of the distribution, where the
x-axis is identical to the main plot. The distributions are consistent
with the stellar parameters derived in Section~\ref{stellar} which
show that the stellar sample is dominated by low-mass stars. For
$\sim$50\% of our stellar sample, the outermost HZ boundary (OHZ$_o$)
lies within 0.5~AU of the star.


\section{Mean Stellar Density and Cluster Dynamics}
\label{dynamics}

In this section, we calculate an estimate for the mean stellar density
of the $\omega$~Cen core. To do this, we adopt the velocity dispersion
data for globular clusters provided by \citet{pry93}. According to
this catalog, $\omega$~Cen has a mean core density of
$\sim$3,000~$M_\odot$/pc$^3$. Based on the stellar distribution
described in Section~\ref{stellar}, the mean stellar mass within the
cluster is $\sim$0.4~$M_\odot$. The average volume occupied by a
single cluster star is thus $1.3 \times 10^{-4}$~pc$^3$ which results
in a mean separation between stars of 0.05~pc (10,000~AU) between
stars. These results are consistent with the estimates of the core
density by \citet{mer97} using cluster dynamics based on radial
velocity measurements.

Numerous observations of $\omega$~Cen, including radial velocities of
individual cluster members, have been used to study the dynamics of
the core and surrounding regions \citep{mer97,rei06,van06}. Close
encounters between stars can greatly effect the local dynamical
environment \citep{ash04,mal07}, particularly the stability of
planetary systems \citep{mal11} and the potential creation of highly
eccentric planetary orbits \citep{mal09,kan14b}. Here we utilize the
stellar encounter rate methodology discussed by \citet{mal07} to
quantify the potential for close encounters in $\omega$~Cen that may
disrupt planetary systems. Specifically, we use Equation~1 of
\citet{mal07} that describes the timescale for a given star to pass
within a distance $r_\mathrm{min}$ of another star:
\begin{equation}
  \tau_\mathrm{enc} \simeq 3.3\times10^7 \mathrm{\, yr} \left(
  \frac{100 \mathrm{\, pc}^{-3}}{n} \right) \left( \frac{v_\infty}{1
    \mathrm{\, km \, s}^{-1}} \right) \left( \frac{10^3 \mathrm{\,
      AU}}{r_\mathrm{min}} \right) \left( \frac{M_\odot}{m_t} \right)
  \label{encounter}
\end{equation}
where $n$ is the stellar number density, $v_\infty$ is the mean
relative speed of the stars at infinity, and $m_t$ is the total mass
of the stars involved in the encounter. The cluster properties
described above and the mean stellar mass of 0.4~$M_\odot$ result in
calculated values of $n = 7.5 \times 10^5$ and $m_t =
0.8$~$M_\odot$. Using the velocity dispersion measurements of
\citet{pry93}, \citet{rei06}, and \citet{and10}, combined with the FOV
of the {\it HST} observations described in Section~\ref{hst}, we adopt
a relative speed for the stars of $v_\infty = 15$~km\,s$^{-1}$. We
then calculate the encounter timescale using Equation~\ref{encounter}
as a function of $r_\mathrm{min}$.

\begin{figure}
  \includegraphics[angle=270,width=8.2cm]{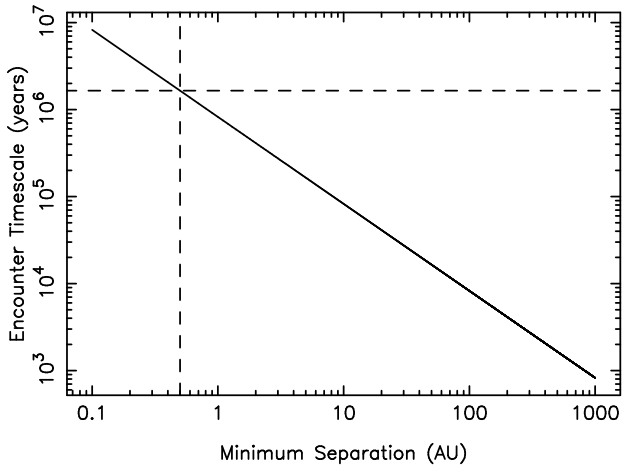}
  \caption{The timescale for a close encounter between two
    0.4~$M_\odot$ stars in the core of $\omega$~Cen as a function of
    the minimum separation of the stars during the encounter (solid
    line). The dashed lines indicate the encounter timescale for a
    minimum separation of 0.5~AU, which encompasses the OHZ outer
    boundary for $\sim$50\% of the observed stars, as described in
    Section~\ref{hz}.}
  \label{encfig}
\end{figure}

The results of these calculations are shown as the solid line in
Figure~\ref{encfig}. The vertical and horizontal dashed lines
represent a minimum stellar separation of $r_\mathrm{min} = 0.5$~AU,
for which the close encounter timescale is $\tau_\mathrm{enc} = 1.65
\times 10^6$~years. The significance of this particular minimum
separation is that it corresponds to the distance from the star within
which the OHZ outer boundary lies for $\sim$50\% of our stellar
sample, as described in Section~\ref{hz}. Notice also that the
encounter timescale for $r_\mathrm{min} = 1,000$~AU is only
$\sim$1,000~years, consistent with the mean distance between stars of
10,000~AU. It is worth noting that these encounter timescale
calculations include only the effect of the MS stars comprised in our
sample, described in Section~\ref{stellar}. The inclusion of the red
giant branch and white dwarf populations, along with other evolved
stars, will have the effect of increasing the total mass of the stars
participating in the encounter. According to Equation~\ref{encounter},
the result of that inclusion would be to decrease the mean time
between close encounters for a given $r_\mathrm{min}$.


\section{Implications for Habitability}
\label{implications}

The search for exoplanets within globular clusters has had a checkered
history, such as the interpretation of 47~Tuc observations described
in Section~\ref{intro}. As such, the prevalence of exoplanets in these
high stellar density environments remains somewhat
uncertain. \citet{dis16} argued that globular clusters are optimal
locations for expansion of advanced civilizations due to the proximity
between stars. The relatively low metallicity of stars in globular
clusters could result in a lower occurrence rate of short-period
jovian planets \citep{fis05}, although more recent studies of Kepler
host star abundances indicate that the occurrence rates of terrestrial
planets and Jupiter analogs are less sensitive to host star
metallicity \citep{buc12,buc18}. A study of open clusters observed
with {\it Kepler} was performed by \citet{cha12} and demonstrated that
planets detected in these environments could be indicative of
perturbations of planetary orbits in globular clusters.

The distribution of HZ boundaries calculated in Section~\ref{hz}
suggests that $\omega$~Cen could be potentially be populated with a
plethora of compact planetary systems that harbor HZ planets close to
the host star. An extreme example of such a system is TRAPPIST-1,
which contains three planets within the HZ of the host star
\citep{gil17}. However, the proximity of the stars combined with the
dynamics of the cluster ensure that close encounters between the stars
are relatively frequent. As shown in Section~\ref{dynamics} and
Figure~\ref{encfig}, a close encounter of $\sim 1$~AU between typical
core cluster members will occur every $\sim 10^6$~years on
average. Even for a minimum encounter separation of $r_\mathrm{min} =
0.01$~AU, comparable to the semi-major axis of the inner planets in
the TRAPPIST-1 system, the timescale for such an event is $\sim
10^9$~years. The result of these frequent disruptive stellar
encounters will be to strip planets from their host stars and create a
large population of free-floating terrestrial planets \citep{mal11}. A
large population of free-floating planets has previously been
constrained from microlensing observations \citep{ban16,cla16} and
predicted from core-accretion theory \citep{ma16}. \citet{hen16}
outlined a strategy through which free-floating planets could be
characterized, including planets within the terrestrial
regime. Furthermore, \citet{ste99} proposed that free-floating planets
with a rich molecular hydrogen atmosphere can retain habitable
conditions at the surface. Thus, despite the dire dynamical
environment of the $\omega$~Cen core, habitable planets in that region
cannot be entirely ruled out.


\section{Conclusions}
\label{conclusions}

The $\omega$~Cen cluster is amongst the most studied objects in the
sky and provides a unique opportunity to study large globular cluster
dynamics as well as the effect on the local group. The {\it HST}
observations of the core have been utilized here to fully explore the
HZ distribution of the stars in that region and we have presented the
first such calculations of HZs in an extremely high stellar density
environment. The peak of the HZ distribution within 0.5~AU of the host
stars is a consequence of the relatively aged population of stars in
the cluster and is a positive aspect of the overall habitability
environment in the $\omega$~Cen core. However, the compact nature of
the HZ regions is more than offset by the potential disruption of
planetary systems, where close encounters of only 0.5~AU are expected
to occur on average every $1.65 \times 10^6$~years. Though the large
resulting population of free-floating terrestrial planets are
intrinsically interesting from formation and dynamical points of view,
the potential for habitability in the $\omega$~Cen core environment is
significantly reduced by such scattering events. The primary lesson
that can be extracted from this analysis is the underlining of the
importance of quantifying the long-term dynamical stability of orbits
inside HZ regions taking into account both internal (planetary)
dynamics and external (stellar) interactions.


\section*{Acknowledgements}

The authors would like to thank Adrienne Cool for providing a wealth
of valuable feedback, and to the anonymous referee, whose comments
greatly improved the quality of the paper. This research has also made
use of the Habitable Zone Gallery at \url{hzgallery.org}, and the NASA
Exoplanet Archive, which is operated by the California Institute of
Technology, under contract with the National Aeronautics and Space
Administration under the Exoplanet Exploration Program. The results
reported herein benefited from collaborations and/or information
exchange within NASA's Nexus for Exoplanet System Science (NExSS)
research coordination network sponsored by NASA's Science Mission
Directorate.


\end{document}